\documentclass[a4paper,11pt]{article}
\usepackage{pos}

\usepackage{multirow} 
\usepackage{subcaption}

\usepackage{physics}
\usepackage[htt]{hyphenat}
\usepackage{soul} 

\usepackage{colortbl}
\definecolor{lightblue}{RGB}{202, 225, 255}
\definecolor{olivegreen}{RGB}{202, 255, 112}
\definecolor{darkolivegreen}{RGB}{85, 107, 47}
\definecolor{littledarkgreen}{RGB}{0,80,0}
\definecolor{firebrick}{RGB}{178,34,34}
\definecolor{darkslateblue}{RGB}{72,61,139}
\definecolor{midnightblue}{RGB}{25,25,112}
\definecolor{darkblue}{RGB}{0,0,139}
\definecolor{indigo}{RGB}{75,0,130}
\definecolor{dodgerblue}{RGB}{30,144,255}
\definecolor{mistyrose}{RGB}{255,228,225}
\definecolor{khaki}{RGB}{240,230,140}
\definecolor{royalblue}{RGB}{65,105,255}
\definecolor{mediumseagreen}{RGB}{60,179,113}
\definecolor{mediumspringgreen}{RGB}{60,179,113}
\definecolor{lime}{RGB}{0,255,0}
\definecolor{limegreen}{RGB}{50,205,50}
\definecolor{teal}{RGB}{0,128,128}
\definecolor{blueviolet}{RGB}{138,43,226}
\definecolor{lightmagenta}{RGB}{255,51,255}
\definecolor{background1}{RGB}{245,255,250}

\graphicspath{{./Figs/}}

\title{Two-link Staggered Quark Smearing in QUDA}
\ShortTitle{Two-link Staggered Quark Smearing in QUDA}

\author[a]{Steven Gottlieb}
\author*[a]{Hwancheol Jeong}
\author[b]{Alexei Strelchenko}

\affiliation[a]{Department of Physics, Indiana University, Bloomington, Indiana 47405, USA}
\affiliation[b]{Scientific Computing Division, Fermi National Accelerator Laboratory, Batavia, Illinois, 60510, USA}

\emailAdd{sg@indiana.edu}
\emailAdd{sonchac@gmail.com}
\emailAdd{astrel@fnal.gov}

\abstract{
  Gauge covariant smearing based on the 3D lattice Laplacian can be used to create extended operators that have better overlap with hadronic ground states. For staggered quarks, we make use of two-link parallel transport to preserve taste properties. We have implemented the procedure in QUDA. We present the performance of this code on the NVIDIA A100 GPUs in Indiana University's Big Red 200 supercomputer and on the AMD MI250X GPUs in Oak Ridge Leadership Computer Facility's (OLCF's) Crusher and discuss its scalability. We also study the performance improvement from using NVSHMEM on OLCF's Summit. Reusing precomputed two-link products for all sources and sinks, it reduces the total smearing time for a baryon correlator measurement by a factor of $100$--$120$ as compared with the original MILC code and reduces the overall time by $60$--$70$\%.
}

\FullConference{%
  The 39th International Symposium on Lattice Field Theory (Lattice2022),\\
  8-13 August, 2022 \\
  Bonn, Germany 
}

\begin{document}
\maketitle

\section{Introduction}
\label{sec:intro}
Lattice QCD calculations require operators that have a strong overlap with particular hadronic states.  For example, 
lattice QCD calculations that study the low energy hadron spectrum benefit from extended operators that have better overlap with the ground state than local operators at a single lattice site. Decomposing a correlation function in terms of energy eigenstates, a two-point correlation function $C (t) = \sum_{\vb{x}} \expval{ \mathcal{O}(t,\vb{x})\, \mathcal{O}^\dagger(0,\vb{0}) }$ can be expressed as
\begin{equation}
  C (t) = \sum_n \left|\mel{0}{\mathcal{O}}{n}\right|^2 e^{- E_n t} \,,
\end{equation}
where $E_n$ is the energy of the $n$-th energy eigenstate. We can extract some low energy properties from it, including the mass from the ground state contribution. If the configuration is gauge-fixed, one can use extended operators without concern about parallel transport; however, by using gauge-covariant smearing of the source one can avoid having to fix the gauge.

There are two popular kinds of gauge covariant smearing: Jacobi smearing and Gaussian smearing. Jacobi smearing is an iterative version of Wuppertal smearing which takes the three-dimensional scalar propagator as a smeared source~\cite{Gusken:1989ad, Gusken:1989qx, Alexandrou:1990dq, UKQCD:1993gym}. The smeared source follows the exponential distribution on a free gauge configuration. Alternatively, Gaussian smearing applies a hopping operator iteratively to the given source. The resulting smeared source follows the Gaussian distribution on a free gauge configuration~\cite{Gusken:1989qx, Alexandrou:1990dq, vonHippel:2013yfa}.

In this paper, we are interested in a variant of Gaussian smearing, which replaces the hopping operator with the three-dimensional lattice Laplacian operator for staggered quarks~\cite{HadronSpectrum:2009krc}. To preserve the taste symmetry of staggered quarks, the Laplacian should extend to the next-to-nearest-neighbor sites. We define the two-link products joining the next-to-nearest-neighbor sites and call this smearing method two-link staggered quark smearing. 

The MILC code with which we started was doing two-site parallel transport by applying single-site parallel transport twice in the same direction.  This requires two communications and two matrix-vector multiplies per direction.
We found that this smearing was taking an inordinate amount of time when done on the CPU. The situation is even worse when other parts of the calculation run on the GPU, because one allocates only one MPI rank per GPU, requiring multi-threading using  OpenMP to use more than one CPU core per rank. Hence, we have implemented the procedure in QUDA~\cite{quda:github, Clark:2009wm, Babich:2011np}.  The exascale computers in the U.S. all make use of GPUs, so more and more of our projects will make use of this timely addition to our codes.

Section \ref{sec:tlSmear} briefly describes the two-link staggered quark smearing. Section \ref{sec:gpu} describes our GPU implementation and algorithmic improvement. In Secs.~\ref{sec:nvidia} and \ref{sec:amd}, we present benchmark results on some (recent or latest) NVIDIA and AMD GPUs, respectively. We also apply our QUDA smearing routine to a baryon correlator measurement in Sec.~\ref{sec:appl} to show how our code performs in a production job. We summarize our conclusions in Sec.~\ref{sec:conc}.

\section{Two-link staggered quark smearing}
\label{sec:tlSmear}
Let us consider a hopping operator $H$ defined by
\begin{equation}
  H(x,y) = \sum_{\mu=1}^3 U_\mu(x)\, \delta_{x+\hat{\mu},\,y} + U_\mu^\dagger (x-\hat{\mu})\, \delta_{x-\hat{\mu},\,y} \,,
\end{equation}
where $x$, $y$ are space-time coordinates and $U_\mu(x)$ is the gauge field. An iterative Gaussian smearing operation to a quark field $\psi(y)$ can be written as
\begin{equation}
  \label{eq:gsmear}
  \widetilde{\psi} = C ( 1 + \alpha H )^n \psi \,,
\end{equation}
where $C, \alpha \in \mathbb{R}$ and $n \in \mathbb{Z}$ are tuning parameters.  For the unit gauge configuration $U_\mu(x)=1$, the smeared field $\widetilde{\psi}(x)$ approaches the Gaussian distribution as the iteration count $n$ grows \cite{Gusken:1989qx, Alexandrou:1990dq, vonHippel:2013yfa}.

The 3D lattice Laplacian $\nabla^2$ is defined by 
\begin{align}
  \nabla^2 (x,y) =& \sum_{\mu=1}^3 \big[ U_\mu(x)\, \delta_{x+\hat{\mu},\,y} + U_\mu^\dagger (x-\hat{\mu})\, \delta_{x-\hat{\mu},\,y} \big] - 6\, \delta_{x,y} \nonumber \\
  =&\, H(x,y) - 6\, \delta_{x,y} \,.
\end{align}
Replacing $U_\mu(x)$ with the \emph{two-link product} $V_\mu(x) \equiv  U_\mu(x) U_\mu(x+\hat{\mu})$ and adjusting the coordinate, we define (ignoring factors of $a$) the two-link 3D lattice Laplacian $\nabla_{\text{two}}^2$: 
\begin{align}
  \label{eq:tlLapl1}
 4 \nabla_{\text{two}}^2 (x,y) &\equiv \sum_{\mu=1}^3 \big[ V_\mu(x)\, \delta_{x+2\hat{\mu},\,y} + V_\mu^\dagger (x-2\hat{\mu})\, \delta_{x-2\hat{\mu},\,y} \big] - 6\, \delta_{x,y} \\
  \label{eq:tlLapl2}
  &= \sum_{\mu=1}^3 \big[ U_\mu(x)\, U_\mu(x+\hat{\mu})\, \delta_{x+2\hat{\mu},\,y} + U_\mu^\dagger (x-\hat{\mu})\, U_\mu^\dagger (x-2\hat{\mu})\, \delta_{x-2\hat{\mu},\,y} \big] - 6\, \delta_{x,y} \,.
\end{align}
Note that $\nabla_{\text{two}}^2$ preserves taste properties for staggered quarks.

\begin{figure}[tb]
  \centering
  \begin{subfigure}[t]{.48\linewidth}
    \includegraphics[width=\linewidth]{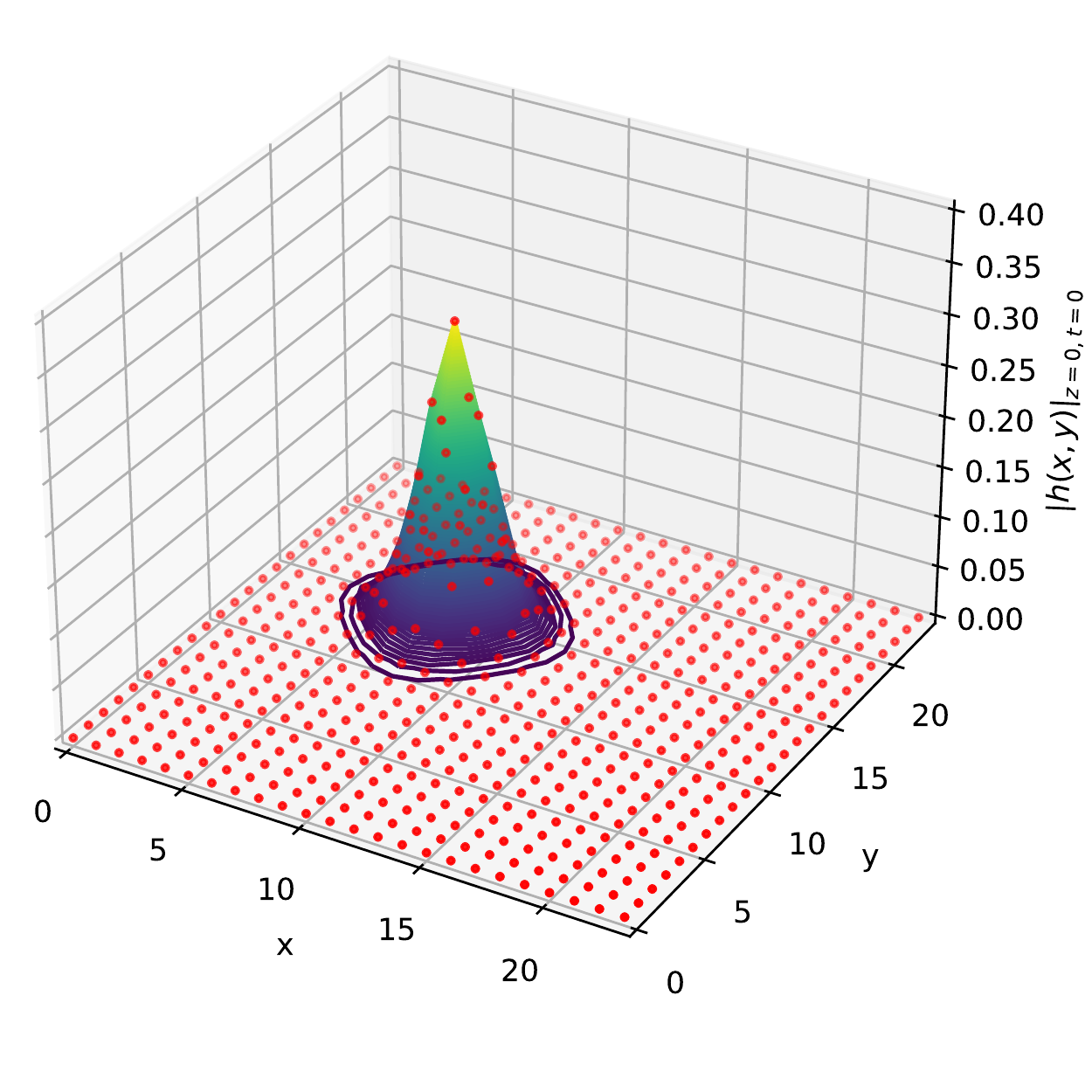}
    \caption{Free gauge ($U = 1$)}
    \label{fig:gsmear_free}
  \end{subfigure}
  \hfill
  \begin{subfigure}[t]{.48\linewidth}
    \includegraphics[width=\linewidth]{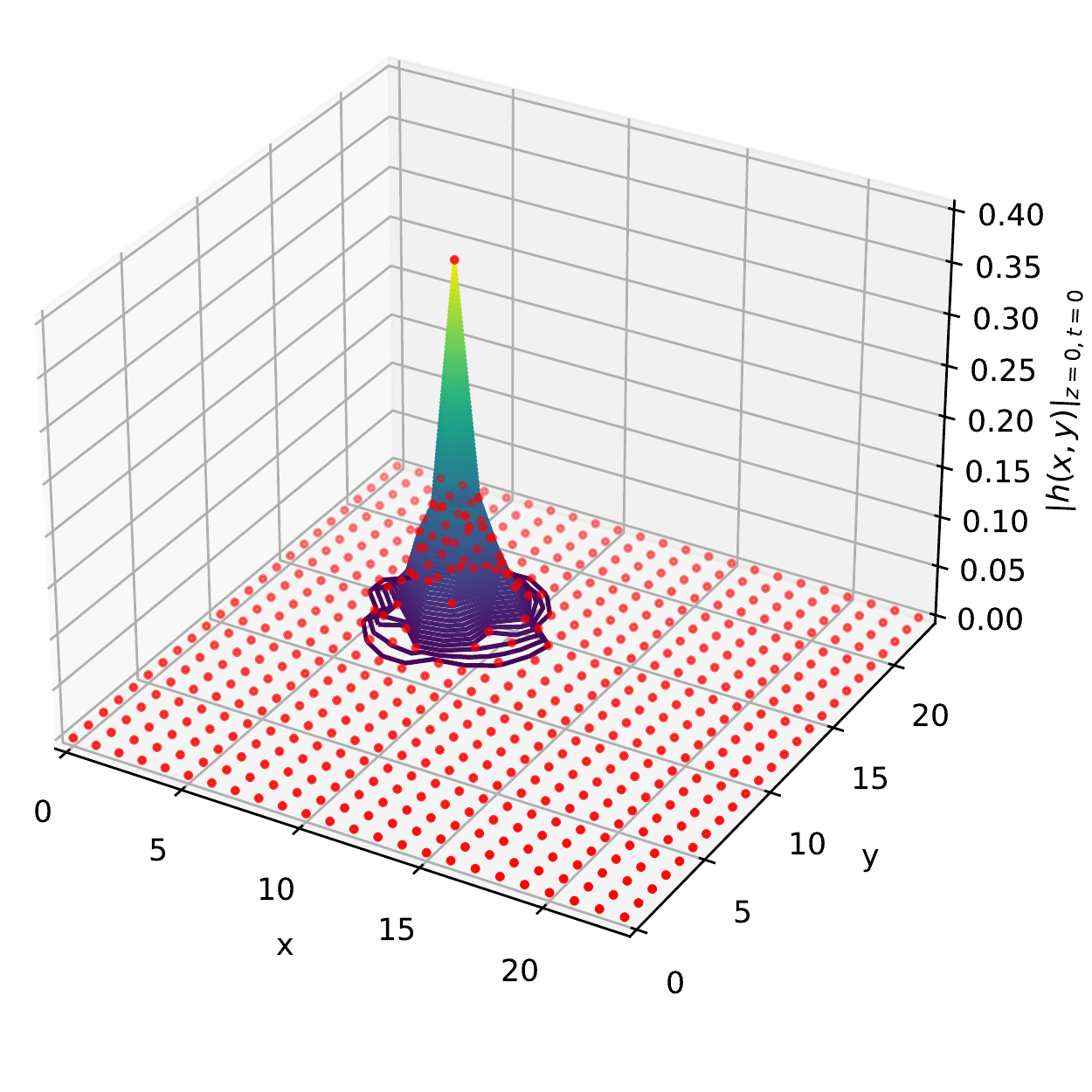}
    \caption{HISQ gauge}
    \label{fig:gsmear_hisq}
  \end{subfigure}
  \caption{Example distributions of a point source smeared by the two-link staggered quark smearing on a free gauge configuration (left) and a HISQ gauge configuration (right). Red data points represent norms of smeared quark field at $(x,y,z,t) = (x,y,0,0)$. Contours are drawn by connecting these points.}
  \label{fig:gsmear}
\end{figure}
If we rewrite Eq.~\eqref{eq:gsmear} in terms of the Laplacian $\nabla^2$,
\begin{equation}
  \widetilde{\psi} = C ( 1 + \alpha ( \nabla^2 + 6 ) )^n \psi \equiv C' ( 1 + \alpha' \nabla^2 )^n \psi \,,
\end{equation}
where $C' \equiv C ( 1 + 6 \alpha )^n$ and $\alpha' \equiv \displaystyle\frac{\alpha}{1+6\alpha}$. Now, we define the two-link staggered quark smearing as
\begin{equation}
  \label{eq:tlSmear}
  \widetilde{\psi} = \left( 1 + \frac{\sigma}{n} \nabla_{\text{two}}^2 \right)^n \psi \,,
\end{equation}
where $\sigma$ and $n$ are tuning parameters. This is a taste-preserving gauge covariant smearing for staggered quarks. Figure \ref{fig:gsmear} shows the distributions of a point source after this smearing is applied on the free gauge configuration (Fig.~\ref{fig:gsmear_free}) and a HISQ gauge configuration (Fig.~\ref{fig:gsmear_hisq}). Although the latter is distorted by the existence of the gauge field, these results imply we can get a better overlap with hadronic ground states with a suitable choice of tuning parameters $\sigma$ and $n$ \cite{Gusken:1989qx, DeGrand:1990dz, UKQCD:1993gym}.
Using a gauge-smeared link in the place of $U_\mu$ can relax the distortion as well as increase the overlap with the ground state \cite{Syritsyn:2009mx}. There are also other approaches that enhance the overlap with some good properties \cite{vonHippel:2013yfa, Bali:2016lva}.

\section{GPU implementation}
\label{sec:gpu}
Before this project began, the MILC code library \cite{milc:github} provided a CPU based routine for the two-link staggered quark smearing.  In need of the GPU equivalent, we have implemented it in QUDA. We have also added an interface for this QUDA smearing in the MILC code, which runs all QUDA benchmarks in this paper.\footnote{We are working on merging the QUDA code and the MILC code interface into the main (develop) branches of QUDA and the MILC code's GitHub repositories, respectively\cite{quda:github, milc:github}.}

We also improved upon the CPU algorithm in the MILC code. Instead of carrying out two consecutive parallel transports (as in Eq.~\eqref{eq:tlLapl2}) at each smearing iteration, we precompute the two-link product $V_\mu$ in advance of smearing, and every iteration just loads it from the memory and uses Eq.~\eqref{eq:tlLapl1}. This two-link product can even be reused for different sources or sinks. In the GPU algorithm, we perform the smearing with significantly less memory traffic, fewer floating point operations, and less communication between MPI ranks. 

\begin{figure}[tb]
  \centering
  \includegraphics[width=0.9\linewidth]{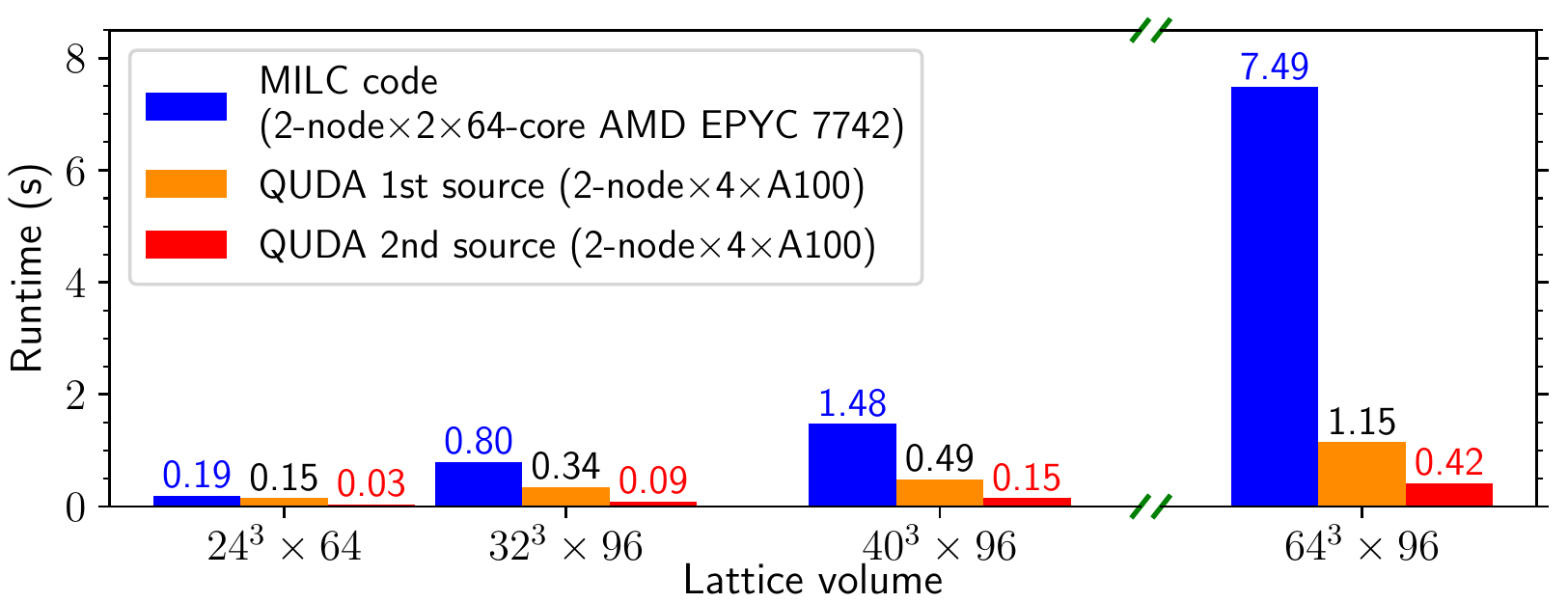}
  \caption{Time taken by smearing a source quark field with $n=50$ two-link staggered quark smearing by the MILC code and QUDA. They are measured on two nodes of Big Red 200. The MILC code is run with one MPI rank per CPU core (total peak double precision FLOPS $\approx$ 14 TFLOPS), and QUDA is run with one MPI rank per GPU (total peak double precision FLOPS $\approx$ 80 TFLOPS). Lattices are divided by $(x,y,z,t)=(8,8,4,1)$ for MILC code runs and $(x,y,z,t)=(2,2,2,1)$ for QUDA runs. Note that splitting in $t$-dimension does not improve the performance of $3$D Gaussian smearing applied to fixed time sources such as point sources or wall sources.
  }
  \label{fig:milc_quda}
\end{figure}
In Fig.~\ref{fig:milc_quda}, we compare the smearing time required by the MILC code to that of our QUDA code using all CPUs and GPUs, respectively, on Big Red 200 at Indiana University. Big Red 200 is similar to Perlmutter at NERSC in that it has a CPU partition in which each node contains two AMD 64-core EPYC 7742 processors and a GPU partition in which each node contains an AMD processor and four NVIDIA A100 GPUs.  Big Red 200 has a Slingshot 10 network, whereas Perlmutter was upgraded from Slingshot 10 to 11.  We ran benchmarks for four different lattice volumes on either two CPU or two GPU nodes. For production jobs, we like to run with a local volume of at least $24^4$ per GPU and $32^4$ or larger is preferred.  Thus, these tests with a fixed number of nodes were not designed for maximum efficiency. In the case of QUDA runs, we also measure the second source smearing that reuses the precomputed two-link product. The first source smearing includes the computation of the two-link product and $n$ iterations of smearing, while the second source smearing does only $n$ iterations of smearing.  Typical computations make use of multiple sources and sinks on the same gauge configuration, so the second source smearing result is quite pertinent. 
The results show that the first source smearing by QUDA on the GPU takes from 0.79 to 0.15 times as long as the MILC code smearing on the CPU, and the second source smearing by QUDA takes from 0.16 to 0.06 times as long as the MILC code. The improvement increases for larger lattice volumes, as the smaller cases are too small for the GPU code to run efficiently.

In this section, we have compared the maximum --- from  the point of view of using all available CPU/GPU resources --- performances of the MILC code smearing and the QUDA smearing on two nodes of a computer with both CPU and GPU nodes. In practice, however, our primary goal is to implement the QUDA smearing in measurements running on GPU nodes. In this situation, smearing running on the CPU can become a severe bottleneck. Section \ref{sec:appl} discusses an example.

\section{Performance on NVDIA GPU}
\label{sec:nvidia}
In this section, we report the performance of the two-link staggered quark smearing in QUDA (QUDA two-link smearing) on two NVIDIA GPU based systems. We smear three different color wall sources in succession. Only the first color source smearing computes the two-link product, while the others reuse it.

\begin{figure}[tb]
  \centering
  \begin{subfigure}[t]{.44\linewidth}
    \includegraphics[width=\linewidth]{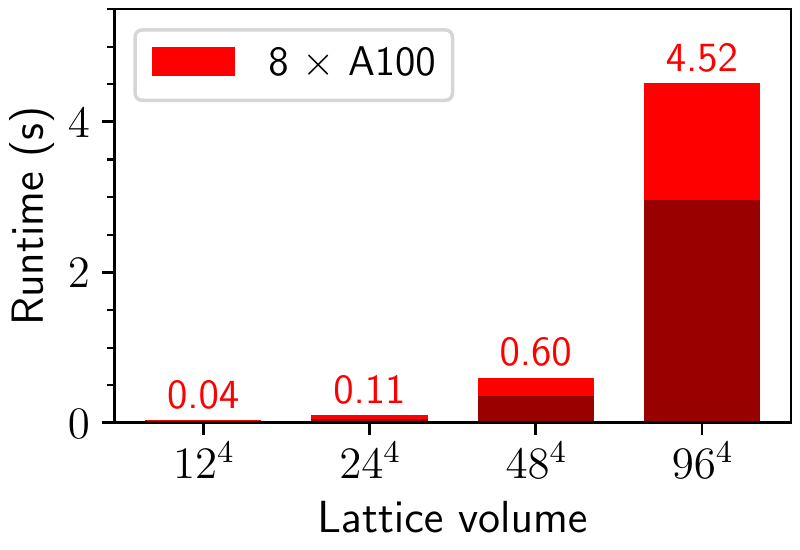}
    \caption{Volume scalability}
    \label{fig:nvidia_volume}
  \end{subfigure}
  \hfill
  \begin{subfigure}[t]{.44\linewidth}
    \includegraphics[width=\linewidth]{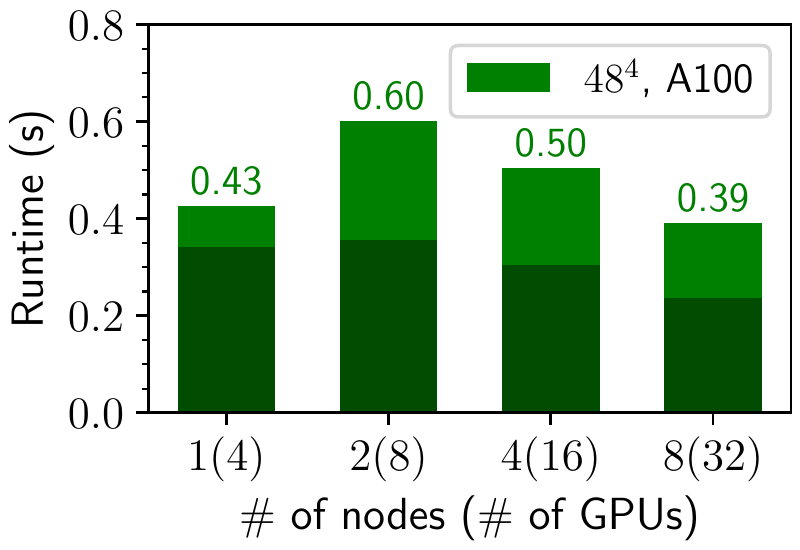}
    \caption{Strong scalability}
    \label{fig:nvidia_strong}
  \end{subfigure}
  \caption{Time taken by smearing three color sources with $n=50$ QUDA two-link smearing on NVIDIA A100 GPUs in Big Red 200. The unshaded region represents the time taken by the two-link computation, and the shaded region represents the time taken by $3 \times n$ iterations of smearing. Lattices are divided by $(x,y,z,t)=(2,2,1,1)$ for $4$ GPUs, $(2,2,2,1)$ for $8$ GPUs, $(4,2,2,1)$ for $16$ GPUs, and $(4,4,2,1)$ for $32$ GPUs.}
  \label{fig:nvidia}
\end{figure}
In Fig.~\ref{fig:nvidia}, we measure the performance and scalability of the QUDA two-link smearing on NVIDIA A100 GPUs in Big Red 200. The total runtime includes one run of two-link computation (unshaded) and three $n=50$ smearing iterations (shaded). We find that the single two-link computation takes around $10$--$40$\% of the time. This indicates the advantage of reusing the two-link product.

Figure \ref{fig:nvidia_volume} presents the smearing time by varying the lattice volume. While the lattice volume increases geometrically by a factor of 16, the total smearing time increases by factors of $2.75$, $5.45$, and $7.53$, respectively. This indicates the computation has not been saturated yet up to the largest lattice volume, so its performance (FLOPS) would be better for a bigger lattice. Still, for small lattices, it would be a fraction of time compared to typical lattice simulation scales. Figure \ref{fig:nvidia_strong} presents the smearing time by varying the number of nodes and GPUs. Note that the two-node run is slower than the one-node run. Even the two-link computation taking three times longer. This implies the communication between off-nodes affects the performance significantly, and it is more prominent for the two-link computation. Excluding the one-node result, the performance improves as the number of nodes increases, but it scales very poorly. Figure \ref{fig:nvidia_volume} and \ref{fig:nvidia_strong} imply that this routine is a communication-intensive calculation.

\begin{figure}[tb]
  \centering
  \includegraphics[width=0.9\linewidth]{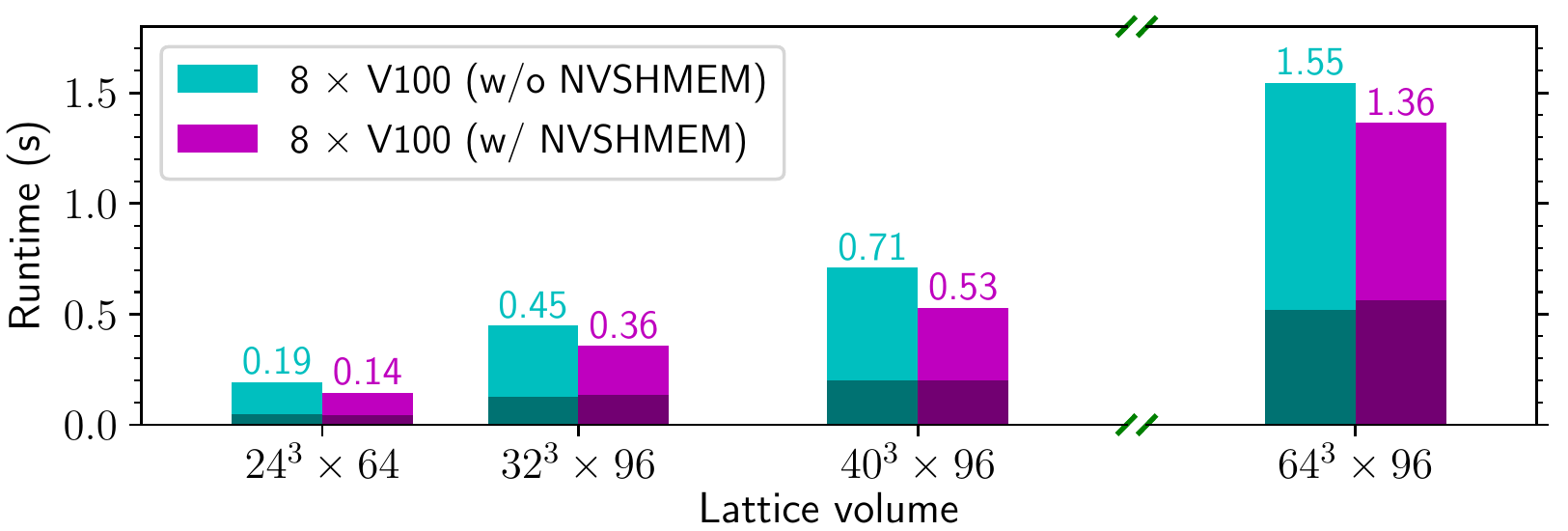}
  \caption{Time taken by smearing three color sources with $n=50$ QUDA two-link smearing on NVIDIA V100 GPUs in Summit with/without NVSHMEM support enabled in QUDA. The unshaded (shaded) region represents the time taken by the two-link computation (smearing iterations). Here, we use two Summit nodes, but only four V100 GPUs out of six per node, because six is not suitable for dividing the spatial dimension of some representative lattices. All lattices are divided by $(x,y,z,t)=(2,2,2,1)$.}
  \label{fig:nvshmem}
\end{figure}
The observation above suggests that the QUDA two-link smearing may perform better with a faster communication environment. Summit at OLCF supports NVIDIA NVSHMEM technology. NVSHMEM improves strong scaling of GPU operations by enabling direct communication between GPUs with shared memory space \cite{nvshmem}. QUDA supports NVSHMEM \cite{quda:nvshmem}. Figure \ref{fig:nvshmem} shows the performance improvement of the QUDA two-link smearing on Summit by enabling NVSHMEM. Here, the two-link computation takes more than half of the total runtime. We find that NVSHMEM reduces the two-link computation time by around $30$--$50$\%.

\section{Performance on AMD GPU}
\label{sec:amd}
QUDA also supports HIP on AMD ROCm platform \cite{quda:hip}, allowing it to run on AMD GPUs. In this section, we report the performance of the QUDA two-link smearing on Crusher at OLCF, an AMD GPU-based system containing hardware identical to that on Frontier. As in Sec.~\ref{sec:nvidia}, we smear three different color wall sources in order, where the two-link product is computed only at the first smearing and reused for others.

\begin{figure}[tb]
  \centering
  \begin{subfigure}[t]{.44\linewidth}
    \includegraphics[width=\linewidth]{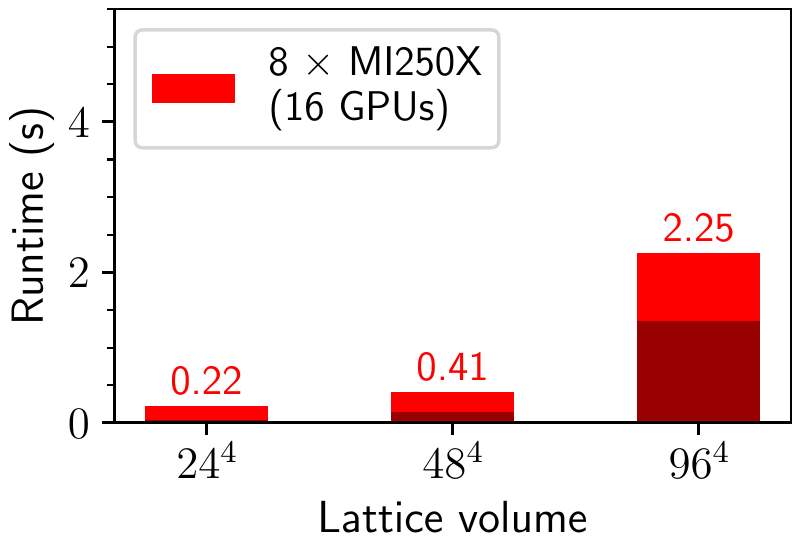}
    \caption{Volume scalability}
    \label{fig:amd_volume}
  \end{subfigure}
  \hfill
  \begin{subfigure}[t]{.44\linewidth}
    \includegraphics[width=\linewidth]{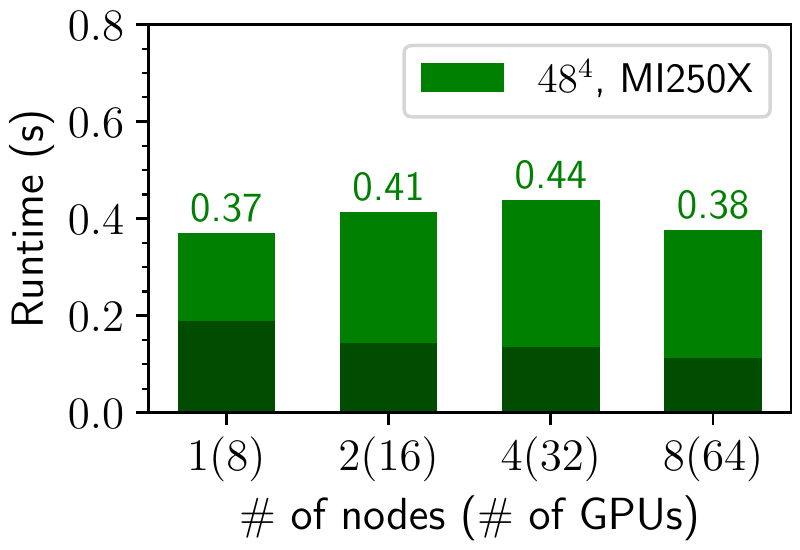}
    \caption{Strong scalability}
    \label{fig:amd_strong}
  \end{subfigure}
  \caption{Time taken by smearing three color sources with $n=50$ QUDA two-link smearing on AMD MI250X GPUs on Crusher. Note that each MI250X contains two GCDs, so the actual number of GPUs used is twice of the number of MI250Xs. The unshaded (shaded) region represents the time taken by the two-link computation (smearing iterations). Lattices are divided by $(x,y,z,t)=(2,2,2,1)$ for $8$ GPUs, $(4,2,2,1)$ for $16$ GPUs, $(4,4,2,1)$ for $32$ GPUs, and $(4,4,4,1)$ for $64$ GPUs.}
  \label{fig:amd}
\end{figure}
\begin{table}
  \centering
  \renewcommand{\arraystretch}{1.1}
  \begin{tabular}{ >{\centering\arraybackslash}m{0.15\linewidth} | >{\centering\arraybackslash}m{0.38\linewidth} | >{\centering\arraybackslash}m{0.38\linewidth} }
    \hline
    & Big Red 200 & Crusher \\
    \hline\hline
    \multirow{2}{*}{GPU} & 4 $\times$ NVIDIA A100 & 4 $\times$ AMD MI250X (=8 GPUs) \\
    & ($\approx$ 40 TFLOPS) & ($\approx$ 210 TFLOPS) \\
    \hline
    GPU & \multirow{2}{*}{1555 GB/s} & \multirow{2}{*}{3277 GB/s} \\    
    Mem. BW & & \\    
    \hline
    \multirow{2}{*}{NIC} & 2 HPE $\times$ Slingshot-10 & 4 $\times$ HPE Slingshot-11 \\
    & (200 Gbps) & (800 Gbps) \\
    \hline
  \end{tabular}
  \caption{GPU and NIC specification of Big Red 200 and Crusher \cite{bigred200, nvidia:a100, crusher, amd:mi250x}. FLOPS numbers represent the double precision peak performance.}
  \label{tab:hardware}
\end{table}
In Fig.~\ref{fig:amd}, we measure the performance and scalability of the QUDA two-link smearing on the AMD MI250X GPUs in Crusher in the same manner as in Fig.~\ref{fig:nvidia}.\footnote{These plots are updated from the ones we presented in the poster, where we observed an abnormal slowdown in the two-link computation on AMD GPUs. It turned out the culprit was a HIP API which was not directly related to the two-link computation. We found a way to avoid this problem and reran the benchmark. We are also working on resolving the issue.} Figure \ref{fig:amd_volume} and \ref{fig:amd_strong} plot the volume and strong scalabilities, respectively. The y-axis ranges are the same as in Fig.~\ref{fig:nvidia_volume} and \ref{fig:nvidia_strong} for easy comparison. The smearing iteration performs approximately twice faster here with MI250X compared to that with A100. This result agrees with our expectation that the bottleneck of this calculation is the GPU memory bandwidth because its compute-to-communication ratio is similar to a typical dslash routine (see Eq.~\eqref{eq:tlLapl1}), and a MI250X has twice faster memory bandwidth than A100 (see Table \ref{tab:hardware}). On the other hand, the two-link computation performs similarly or slower on the MI250X compared to the A100. The only exception is the $96^4$ lattice result, where we observe the expected twice faster performance. Regarding the scalability, we observe a poor scaling as we observed from the A100 GPU.

\section{Application: Baryon correlator measurement}
\label{sec:appl}
\begin{figure}[tb]
  \centering
  \includegraphics[width=0.9\linewidth]{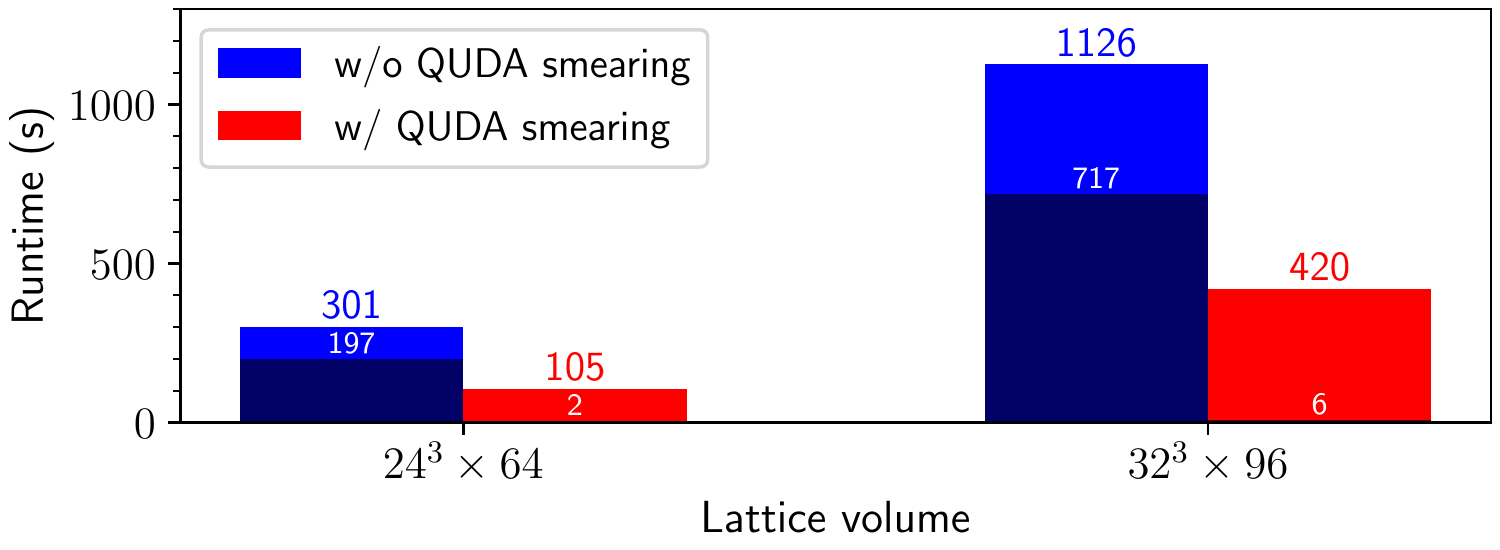}
  \caption{Total time taken by a baryon correlator measurement employing $72$ source/sinks which are smeared by the $n=30$ QUDA two-link smearing. The measurement is carried out on two nodes of Big Red 200's GPU partition using four CPU cores and GPUs per node. With or without the QUDA two-link smearing enabled, all other QUDA-supported calculations are performed on the GPU. The shaded region represents the total smearing time.}
  \label{fig:baryon}
\end{figure}
Our baryon correlator calculations include many sources and sinks. Figure \ref{fig:baryon} shows the total runtime taken by an example baryon correlator measurement. Most parts of the calculation were already implemented in QUDA. With the QUDA smearing disabled, the smearing is carried out on the CPU by the MILC code smearing routine while other major calculations are carried out on the GPU by QUDA. Note that here we use only the same number of CPUs as of GPUs. The result shows that the source/sink smearing takes up around $60$--$70$\% of the total measurement time when it is run on the CPU. However, enabling the QUDA smearing reduces the smearing time by a factor of $100$--$120$, requiring only around $1$--$2$\% of the total measurement time.  Thus, the total time for the job is reduced to about 30--40\% of the original time.

\section{Conclusion}
\label{sec:conc}
We have implemented two-link staggered quark smearing in QUDA. When run on eight NVIDIA A100 GPUs, it performs up to six times faster than the MILC code run on total 256 cores of four AMD EPYC 7742 CPUs. 
Reusing the precomputed two-link product for another source on the same gauge field increases the difference up to 18 times.

The scaling behavior of this routine on both NVIDIA A100 and AMD MI250X is rather poor, especially for the two-link computation part. NVSHMEM can improve the performance of the two-link computation by $30$--$50$\%. For a baryon correlator measurement, reusing the two-link product for all $72$ sources/sinks reduces the total smearing time from $60$--$70$\% to $1$--$2$\% of the total measurement time. Thus, even without further optimization, this code can be useful for GPU jobs that require many smearings for sources or sinks.

\acknowledgments{
  %
  This research was supported by the Exascale Computing Project (17-SC-20-SC), a collaborative effort of the U.S. Department of Energy Office of Science and the National Nuclear Security Administration.
  We gratefully acknowledge support by the U.S. Department of Energy, Office of Science under award DE-SC0010120.
  This research was supported in part by Lilly Endowment, Inc., through its support for the Indiana University Pervasive Technology Institute.
  This research used resources of the Oak Ridge Leadership Computing Facility at the Oak Ridge National Laboratory, which is supported by the Office of Science of the U.S. Department of Energy under Contract No. DE-AC05-00OR22725.
  We thank the QUDA~\cite{Clark:2009wm, Babich:2011np} developers whose names can be found at their website\cite{quda:github}~.

}

\bibliographystyle{JHEP-hc}
\bibliography{ref}

\end{document}